\documentclass[aip,reprint]{revtex4-1}

\usepackage{color, graphicx}
\usepackage{dcolumn}
\usepackage{bm}
\usepackage{amssymb}
\usepackage{latexsym}
\usepackage{amsfonts}
\usepackage{amsmath}
\usepackage{multirow}
\usepackage{ifthen}

\draft 

\begin{document}


\title{Tunable magneto-granular phononic crystals} 



\author{F. Allein}
\author{V. Tournat}
\author{V. E. Gusev}
\author{G. Theocharis}
\affiliation{LAUM, UMR CNRS 6613, Universit\'e du Maine, Avenue O. Messiaen, 72085 Le Mans, France}


\date{\today}

\begin{abstract}
This paper reports on the study of the dynamics of 1D magneto-granular phononic crystals composed of a chain of spherical steel beads inside a properly designed magnetic field. This field is induced by an array of permanent magnets, located in a holder at a given distance from the chain. The theoretical and experimental results of the band gap structure are displayed, including all six degrees of freedom for the beads, i.e. three translations and three rotations. Experimental evidence of transverse-rotational modes of propagation is presented; moreover, by changing the strength of the magnetic field, the dynamic response of the granular chain is tuned. 
The combination of non-contact tunability with the potentially strong nonlinear behavior of granular systems ensures the suitability of magneto-granular phononic crystals as nonlinear, tunable mechanical metamaterials for use in controlling elastic wave propagation. 

\end{abstract}


\maketitle 


The ability to control the propagation of elastic waves has been widely investigated in phononic crystals, which are a class of engineered media composed of periodic arrays of scattering inclusions in a homogeneous host material \cite{phononic}.
The propagation of sound in phononic crystals is driven by the interference between Bragg scattered waves. Similar to the electronic band structure of semiconductors and the electromagnetic band structure of photonic crystals, the phononic crystal structure does not allow certain frequency ranges to propagate within it. The existence of these forbidden bands makes phononic crystals suitable for direct applications, such as mechanical frequency filtering and sound insulation. In addition, the removal of inclusions along some pathways produces acoustic waveguides, demultiplexers and other elastic wave devices. To foresee the next generation of elastic devices, it would be necessary to introduce a certain degree of frequency tunability in the phononic properties. Greater interest has been shown along these lines over the last few years, and many solutions have been proposed by a number of authors; these include geometric changes to the structure by applying external stresses \cite{stress} and changes to the elastic characteristics of the constitutive materials through application of external stimuli, like an electric field \cite{electric}, temperature changes \cite{temperature} or a magnetic field \cite{magnetic}. 

Among the various proposed phononic crystals, granular crystals \cite{chapterG}, which are closely-packed ordered arrangements of elastic particles
(mainly spheres) in contact, have been the topic of a large body of recent work. Due to the Hertzian contact interaction between particles \cite{Hertzbook}, the dynamic vibration response of these structures may be nonlinear and tunable. These features make granular chains a perfect medium for studying fundamental wave structures, including solitary waves with a highly localized waveform \cite{compacton,gr1} or discrete breathers \cite{gr4}, as well as in engineering applications, e.g. tunable vibration filters \cite{tunable}, impulse energy protectors \cite{gr5}, acoustic lenses \cite{Spadoni}, acoustic rectifiers \cite{Nature11}, tunable functional switches \cite{Gonella}.

Provided the applied frequency range remains considerably less than the individual bead resonances,
the phononic band structure of the granular crystals can be obtained by means of discrete lattice models, in which the contact deformations are modeled by springs. Due to the finite size of particles and the friction between them, attention must be paid not only to the normal contact stiffness but also to the tangential and torsional contact stiffnesses, thus suggesting that rotational degrees of freedom be included in the analysis. Coupled rotational-translational elastic waves were observed in three-dimensional, hexagonal closely-packed granular crystals \cite{PRL_Aurel}, the dispersion relation of which is decribed by a three-dimensional discrete lattice model that includes the rotational degrees of freedom. Moreover, a two-dimensional discrete lattice model, with particles possessing one translational and two rotational degrees of freedom, has been applied to a monolayer granular phononic membrane \cite{NJP_VT}. The significance of micro rotations has also been recently revealed experimentally in colloidal-based metamaterials \cite{Nick} as well as in torsional waves within granular chains \cite{PRL_Torsional}.

\begin{figure}[ht!]
\includegraphics[width=8.5cm]{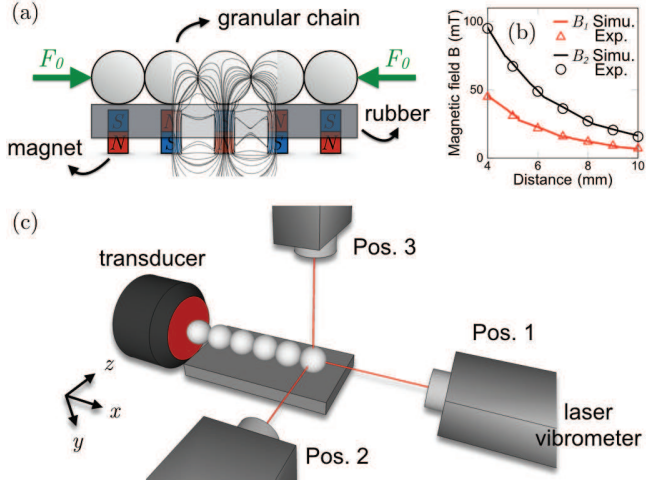}
\caption{(a) Schematic diagram of the granular chain on top of permanent magnets within a rubber. Forces and magnetic flux lines induced by the magnetic field have been superimposed. (b) Magnetic flux density (mTesla) vs. distance from the upper surface of the magnet for $B_1 = 1.32$~T (red) and $B_2 = 1.37$~T (black). (c) Schematic diagram of the experimental set-up. This configuration allows us to drive the first bead of the chain along the $x$, $y$ or $z$ direction by changing the position of the transducer. The laser vibrometer, which is sensitive to changes in the optical path length, enables detecting the motion of the reflecting surface along just the optical beam direction. The vibrometer has been placed in 3 different positions (i.e. Pos. 1, 2 and 3) in order to detect displacements along the $x$, $z$ and $y$ axes, respectively.}\label{fig_beadmagnet}
\end{figure}

This work is devoted to designing and studying tunable magneto-granular phononic crystals. Tunability is demonstrated by introducing granular chains made of ferromagnetic materials and then applying an external magnetic field. Since ferromagnetic materials can be strongly magnetized in the presence of a magnetic field, changing the magnetic field strength allows tuning the interparticle forces and consequently, the dynamic response of the granular crystal.
The proposed magneto-granular crystal is shown in Fig.~\ref{fig_beadmagnet}(a); it is composed of a chain of spherical stainless steel beads placed on top of a rubber substrate, in which permanent neodymium magnets (NdFeB) are laid out in a linear array. This set-up offers the advantages of straightforward construction and non-contact tunability via external magnetic fields. 
The use of magnetic elements to control wave propagation in discrete systems has been explored in various works \cite{magnets}. 
Soft rubber has been chosen as a substrate in order to ensure the high impedance contrast between the granular chain and the substrate.
This contrast serves to reduce their mechanical coupling while the remaining energy leakage is highly attenuated due to the high viscoelasticity of rubber. In addition, the effective stiffness originating from contact deformation between the spheres and the substrate is very small relative to the contact stiffnesses between spheres. It can, thus be assumed that the granular chain is free, in ignoring coupling with the substrate.

 
The interaction force between soft magnetic particles in the presence of uniform or non-uniform applied magnetic fields was investigated in \cite{fujita,rahmatian}. In the present case, due to the strong non-uniformity of the magnetic field, analytical formulae for the magnetic forces between spheres are not available. As a result, numerical simulations can be run to calculate the magnetic flux density of our configuration, while experimental measurements yield estimates of the attractive forces between two adjacent spheres. In Fig. \ref{fig_beadmagnet}(a), the magnetic flux lines have been superimposed on the schematic diagram of the set-up.  These lines have been calculated (using FEM simulations) in considering spheres of diameter $d=15.875$ mm, made of stainless steel (Young's modulus $E=200$ GPa, Poisson's ratio $\nu=0.3$) and permanent cylindrical NdFeB magnets with remanent magnetization $B_2= 1.37$~T at a distance $D=10$~mm from the center of the spheres, as measured from the upper surface of the cylindrical magnet. A large concentration of lines are visible at the contacts, which indicates the attractive normal force between adjacent spheres and between spheres and substrate. In Fig.~\ref{fig_beadmagnet}(b), the simulated (by FEM analysis) and measured magnetic field (using a Gauss-meter \emph{F.W. Bell 5180}) are plotted vs. distance from the upper surface of the cylindrical magnets of $B_1 = 1.32$~T and $B_2 = 1.37$~T. It can be observed that major variations in magnetic field strength are possible by either placing the permanent magnets at different distances or changing their remanent magnetization. For example, in using the $B_1$ permanent magnets, the magnetic field at $D=10$~mm from the upper surface of the magnet is reduced by 42\% compared to $B_2$ permanent magnets, see Fig.~\ref{fig_beadmagnet}(b). This effect influences the strength of the static normal forces between spheres $F_0$, see Fig.~\ref{fig_beadmagnet}(a), induced by the external magnetic field. Such an effect can actually be estimated by measuring the pulling force required to separate two adjacent particles when using a dynamometer. From an experimental standpoint, we measured $F_0 = 0.3 \pm 0.05$~N and $F_0 = 1 \pm 0.15$~N for the case of permanent magnets of $B_{1}=1.32$~T and $B_{2}=1.37$~T, respectively.

\begin{figure}[ht!]
\includegraphics[width=8.5cm]{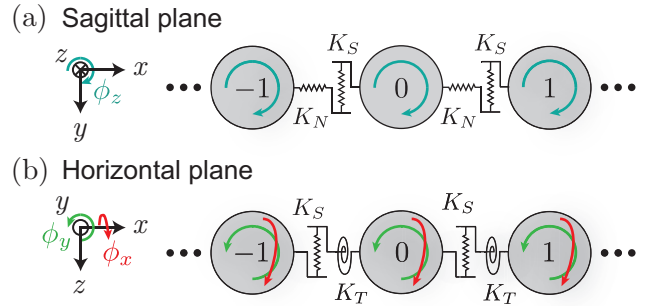}
\caption{Representation of the infinite free granular chain for: (a) displacement and rotation motion in the sagittal plane ($x,y$), and (b) in the horizontal plane ($x,z$).} \label{fig_Shematic_chain_theory} 
\end{figure}

To describe the three-dimensional dynamics of a granular chain, let's consider the mass-spring model shown in Fig.~\ref{fig_Shematic_chain_theory}.
For a better representation and understanding, the possible motions are presented in two distinct planes, called "sagittal" for the ($x,y$) plane and "horizontal" for the ($x,z$) plane. 
This model considers all $6$ degrees of freedom for each sphere of the chain: $3$ displacements along the $x,y,z$-axis, and $3$ rotations around the same axes. 
Between two adjacent spheres, let's consider normal, shear and torsional coupling to be characterized by constant stiffnesses $K_N$, $K_S$ and $K_T$, respectively.
For the case of stiff and macroscopic elastic spheres, $K_N$, $K_S$ and $K_T$ are calculated by means of Hertzian contact mechanics and expressed as follows:
\begin{eqnarray}
K_N  & = & \left( \dfrac{3R}{4}F_0 \right)^{1/3}{E}^{2/3}(1-{\nu}^2)^{-2/3},\\
K_S & = &  (6F_0 R)^{1/3} {E}^{2/3} \dfrac{(1-{\nu}^2)^{1/3}}{(2-\nu)(1+\nu)}, \\
K_T  & = &  \left( 2 R \right) \left( 1 - \nu\right) F_0.
\label{stiffness}
  \end{eqnarray}

By assuming small displacements and rotations, the motion equations of the zeroth particle are given by:

\begin{subequations}
\begin{eqnarray}
m \ddot{u}_{0,x} & = & K_N \left( u_{1,x} - 2 u_{0,x}  + u_{-1,x}  \right), \\
m \ddot{u}_{0,y} &  = & K_S \left( u_{1,y} - 2 u_{0,y}  + u_{-1,y} \right) \\
        & &  + K_S R \left( \phi_{-1,z} - \phi_{1,z} \right) ,  \label{subeq_explmy}  \nonumber\\
m \ddot{u}_{0,z} &  = & K_S \left( u_{1,z} - 2 u_{0,z}  + u_{-1,z} \right)  \\
& & + K_S R \left( \phi_{-1,y} - \phi_{1,y} \right),  \label{subeq_explmz}\nonumber\\
 I \ddot{\phi}_{0,x} &  = & K_T \left( \phi_{1,x} - 2 \phi_{0,x}  + \phi_{-1,x}  \right),\label{subeq_explpsix} \\
 I \ddot{\phi}_{0,y} &  = & -K_S R^2 \left( \phi_{1,y} + 2\phi_{0,y} + \phi_{-1,y} \right) \\
&&  + K_S R \left(u_{1,z} - u_{-1,z} \right), \nonumber\label{subeq_explpsiy}
\nonumber \\
I \ddot{\phi}_{0,z} &  = & -K_S R^2 \left( \phi_{1,z} + 2\phi_{0,z} + \phi_{-1,z} \right) \\
&&  + K_S R \left(u_{1,y} - u_{-1,y} \right), \nonumber\label{subeq_explphiz}
\end{eqnarray}
\label{eq_setmotion}
\end{subequations}
where $m$ is the mass, $R$ the radius, and $I$ the moment of inertia of the particle (for homogeneous spheres, $I = \frac{2}{5}mR^2$), $u_{n,i}$ and $\phi_{n,i}$ denote respectively the displacement and rotation of bead $n$ along the $i = x,y,z$-axis.

The solutions to Eqs.~(\ref{eq_setmotion}) are plane waves propagating in the $x$-direction in the form of ${\bf V_n} = {\bf v}e^{i\omega t - i k_x x_n}$, where $k_x$ is the complex wave number, ${\bf v}$ the amplitude vector and $\omega$ the angular frequency. Substituting these solutions into the set of Eqs.~(\ref{eq_setmotion}) results in the eigenvalue problem
\begin{eqnarray}
 {\bf S} {\bf v} = -\Omega^2 {\bf v}, 
 \label{eq_eigen}
  \end{eqnarray}
where $\Omega = \omega / \omega_0$ is the reduced frequency with $\omega_0 = \sqrt{K_S/ m}$ and S is a $6\times6$ dynamic matrix \cite{Sm}. The solution to this eigenvalue problem yields the $\Omega(k)$ dispersion relation. 

\begin{figure}[ht!]
\includegraphics{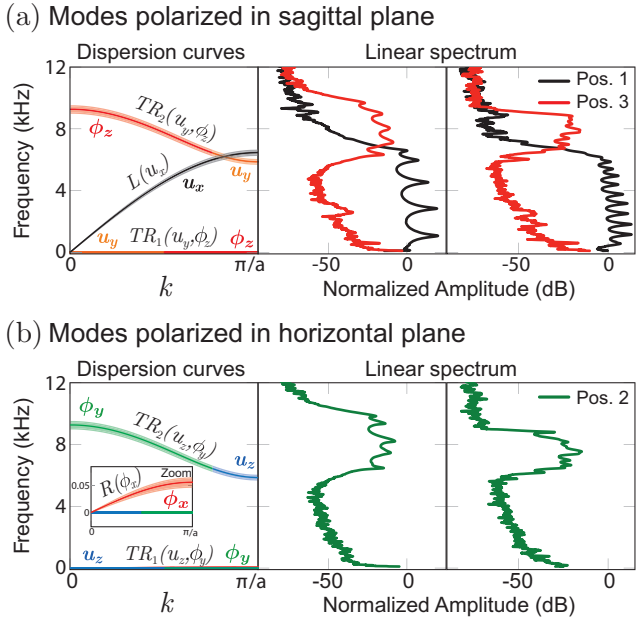}
\caption{Dispersion curves (left) for the infinite free granular chain when considering the experimentally measured $F_0 = 1$~N (solid line) with an error of $\pm 0.15$~N (shadow zone). The linear spectrum of the signal (velocity amplitude) detected by the laser vibrometer for a chain composed of 5 beads (center) and 15 beads (right) (a) in the sagittal plane ($x$,$y$), and (b) in the horizontal plane ($x$,$z$). For both chains, permanent magnets of $B_2$ remanent magnetization have been used. Measurements are performed on the last bead of the chain in the same direction as the excitation polarization. Pos.~1 (black line) measurement along the $x$-axis, Pos.~2 (green line) along the $z$-axis, and Pos.~3 (red line) along the $y$-axis.
}\label{fig_results}
\end{figure}

The sagittal plane (see left panel of Fig.~\ref{fig_results}(a)), contains three branches: one corresponds to the longitudinal mode with displacement just along the $x$-axis, denoted $L (u_x)$, while the other two correspond to transverse-rotational mode, denoted $TR_{1,2}$ ~($u_y,\phi_z$) with displacements along the $y$-axis and rotations along the $z$-axis. $TR_1$ is a zero-frequency branch resulting from the counterbalance between rotational and transverse motions; further details on zero-frequency TR branches can be found in \cite{Johnson, Pichard_PRE2014}. The component predominance in the TR branches is determined by identifying the largest component of the calculated eigenmodes of Eq.~(5), as denoted in Fig.~\ref{fig_results}(a,b). For example, in the lower part of the $TR_2$ branch the predominant component is $u_y$, while it is $\phi_z$ in the upper part. The horizontal plane (see left panel of Fig.~\ref{fig_results}(b)), displays two TR branches, denoted $TR_{1,2} ~(u_z,\phi_y)$, and one branch corresponding to the torsional modes, denoted $R~(\phi_x)$ \cite{PRL_Torsional}. This branch is at very low frequencies due to the small value of the torsional stiffness $K_T$, compared to the effective rotational stiffness $K_S R^2$ provided by shear interactions. $TR_1$ is also a zero-frequency branch due to the counterbalance between rotational and transverse motions, while $TR_2~ (u_z,\phi_y)$ is similar to the $TR_2~ (u_y,\phi_z)$ of the sagittal plane. This result stems from the symmetry of the rotational $\phi_z, \phi_y$ and transverse $u_y,u_z$ motions of the free-standing granular chain (see Fig.~\ref{fig_Shematic_chain_theory}).

Let's now examine the experimental work conducted  to verify the theoretical dispersion relation. Two types of transducers (a longitudinal \emph{Panametrics V3052} and a shear \emph{Panametrics V1548}) have been introduced to drive the chain in three directions ($x,y,z$). A bead is glued to each of the transducers in order to ensure the same contact between the driver and the first bead of the chain. The excitation signal is a sweep sine with a $65$ s duration and a range from $100$~Hz to $15$~kHz. Particle velocity are measured by a laser vibrometer \emph{Polytec OFV-503} with a sensitivity of $5$~mm/s/V. This vibrometer have been placed in three different positions to detect all three translational motions of the bead. This experimental set-up is presented in Fig.~\ref{fig_beadmagnet}(c). The granular chain is composed of $5$ or $15$ stainless steel beads of diameter $d=15.875$ mm. The permanent magnets are NdFeB cylinders of remanent magnetization $B_2=1.37~T$, located 10 mm from the center of the spheres. For this case, as mentioned above, the measured static normal force is $F_0=1$~N. Eqs.~(1)-(3) therefore yield $K_N = 6.60 \cdot10^{6}$~N/m, $K_S= 5.44 \cdot 10^{6}$~N/m, $K_T= 1.12 \cdot 10^{-2}$~m$\cdot$N and $K_S R^2 = 3.42 \cdot 10^{2}$~m$\cdot$N~$\approx 3 \times 10^4~K_T$.

Fig.~\ref{fig_results} plots  the theoretical dispersion curves for both plane (sagittal and horizontal), along with the experimentally derived linear spectra. 
From a direct comparison, a very good agreement can be observed between theoretical and experimental findings. In addition, the appearance of the upper branches in the experiments provides evidence of the existence of coupled transverse/rotational branches. The torsional branch $R ~(\phi_x)$ cannot be detected by the laser vibrometer. 
In the case of a $5$-bead chain (central panels of Fig.~\ref{fig_results}), the $5$ resonances in each branch are clearly visible. In the case of a 15-bead chain and under longitudinal excitation (right panels of Fig.~\ref{fig_results} - Pos.1), only the first $10$ resonances can be distinguished.
This result offers evidence that increasing the system length strengthens the dissipation effect, especially around the edges of propagating bands.
In addition, at high frequencies under both longitudinal and shear excitation, the corresponding propagating wavelengths are on the order of the lattice period. Hence, a small amount of disorder induced by bead misalignments could lead to a stronger smearing of the relevant resonances.

\begin{figure}[ht!]
\includegraphics{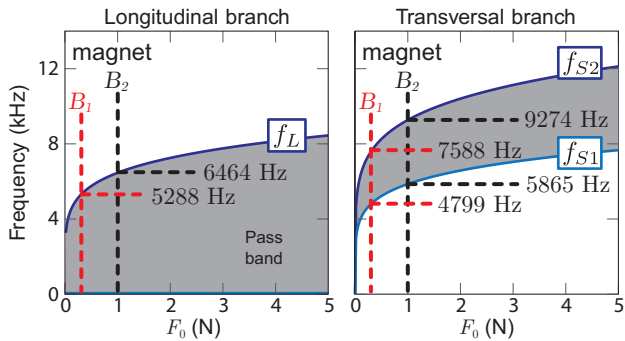}
\caption{Cutoff frequencies for the L (left) and TR (right) branches vs. normal force $F_0$. The experimentally estimated forces are $F_0 = 0.3$~N (dashed red) and $F_0 = 1$~N (dashed black) for the permanent magnets of $B_1$ and $B_2$ remanent magnetization, respectively, with both placed a distance $10$~mm from the center of the beads. The gray zones correspond to the pass bands. The torsional cutoff frequency is not shown.
}\label{fig_Tunability_cutoff_Theory}
\end{figure}

With the validity of the theoretical model now verified, let's continue with a demonstration of the tunability of the magneto-granular crystal by an external magnetic field. Fig.~\ref{fig_Tunability_cutoff_Theory} presents the dependence of cutoff frequencies on the static normal force $F_0$ between beads. The cutoff frequencies are given analytically by:
\begin{eqnarray}
f_{L} & = & \dfrac{1}{2 \pi} \sqrt{\dfrac{4 K_N}{m}}, \label{eq_fL} \\
f_{S1} & = & \dfrac{1}{2 \pi} \sqrt{\dfrac{4 K_S}{m}}, \label{eq_fS1} \\
f_{S2} & = & \dfrac{1}{2 \pi} \sqrt{\dfrac{4 K_S R^2}{I}}.  \label{eq_fS2}
\end{eqnarray}
By changing $F_0$, which in our case is equivalent to changing of the strength of the magnetic field induced by the permanent magnets, the cutoff frequencies of the L and TR branches can be tuned. Fig.~\ref{fig_Tunability_cutoff_Theory} also depicts the expected theoretical cutoff frequencies using the two permanent magnets of $B_1$ and $B_2$, which correspond to $F_0 = 0.3$~N and $F_0 = 1$~N respectively.

\begin{figure}[ht!]
\includegraphics{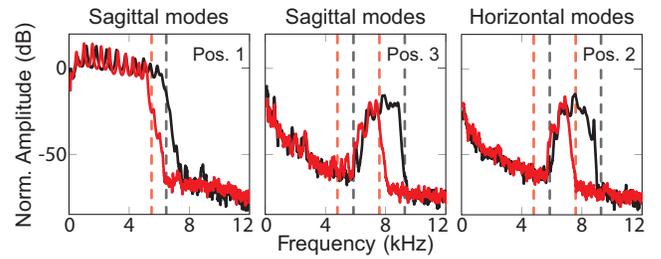}
\caption{Linear spectrum of the velocity signal received for a chain of 15 beads under $B_1$ (red) and $B_2$ (black) permanent magnets at a distance of $10$~mm from the center of the spheres. In the left panel, the dashed lines correspond to analytical cutoff frequencies of the longitudinal branch Eq.~(\ref{eq_fL}) along with the experimentally estimated forces. The dashed line in both the center and right panels shows the expected analytical transverse cutoff frequencies from Eq.~(\ref{eq_fS1}) and~(\ref{eq_fS2}).}\label{fig_Tunability_Exp}
\end{figure}

Fig.~\ref{fig_Tunability_Exp} shows on the same plot the experimental linear spectra of a chain composed of $15$ beads, placed $10$mm above the linear array of permanent NdFeB magnets of $B_1$ (red line) and $B_2$ (black line), respectively. In the former case, introducing the experimentally measured  $F_0 = 0.3$~N into Eqs. (1)-(3) yields  $K_N = 4.42 \cdot10^{6}$~N/m, $K_S= 3.64 \cdot10^{6}$~N/m, $K_T= 3.34 \cdot 10^{-3}$~m$\cdot$N and $K_S R^2= 2.29 \cdot 10^{2}$~m$\cdot$N~$\approx 10^5~K_T$. In Fig.~\ref{fig_Tunability_Exp}, the dashed lines denote the predicted theoretical values of the cutoff frequencies. Band structure tunability is observed by changing the remanent magnetization. More specifically, using the $B_1$ permanent magnets instead of the $B_2$ magnets reveals a downshift of the band edges of roughly~18\%.

In conclusion, this paper has  presented the development and dynamics of a magneto-granular phononic crystal, composed of spherical ferromagnetic spheres in contact, in the presence of a properly designed magnetic field.
This field has been induced by strong permanent magnets arranged in a row inside a rubber holder at a given distance from the granular chain. In considering all degrees of freedom, the full set of dispersion relations have been derived. Experiments have revealed the existence of TR branches in a granular chain that moreover can be tuned by the external magnetic field. The low static load $F_0$ values endow the system with a strong nonlinear response. Preliminary experiments have demonstrated these strong nonlinear dynamics through the presence of higher harmonics and bifurcation instabilities \cite{Lydon}. The combination of the non-contact tunability and strong nonlinear behavior allows envisaging the development of nonlinear, tunable mechanical metamaterials for controlling elastic wave propagation.

We thank A. Merkel for helpful comments. G. T. acknowledges financial support from FP7-CIG (Project 618322 ComGranSol).

\nocite{*}


\begin{thebibliography}{99}

\bibitem{phononic}
Acoustic Metamaterials and Phononic Crystals, edited by P. A.
Deymier (Springer-Verlag, Berlin, Heidelberg, 2013).

\bibitem{stress} 
K. Bertoldi and M. C. Boyce, Phys. Rev. B \textbf{77}, 052105 (2008).
\bibitem{electric} 
J.-Y. Yeh, Physica B \textbf{400}, 137 (2007).
\bibitem{temperature} 
K. L. Jim, C. W. Leung, S. T. Lau, S. H. Choy, and H. L. W. Chan, Appl.
Phys. Lett. \textbf{94}, 193501 (2009).
\bibitem{magnetic}
J.-F. Robillard, O. Bou Matar, J. O. Vasseur, P. A. Deymier, M. Stippinger, 
A.-C. Hladky-Hennion, Y. Pennec, and B. Djafari-Rouhani,  Appl. Phys. Lett. \textbf{95}, 124104 (2009).

\bibitem{chapterG} G. Theocharis, N. Boechler, and C. Daraio, in Acoustic Metamaterials and Phononic Crystals (Springer, New York, 2013), pp. 217-251.

\bibitem{Hertzbook}
 K. L. Johnson,  \textit{Contact Mechanics} (Cambridge Univ. Press, 1985);
V. F.  Nesterenko, \textit{Dynamics of Heterogeneous Materials} (Springer, 2001).

\bibitem{compacton} V. F. Nesterenko, J. Appl. Mech. Tech. Phys. \textbf{24}, 733 (1984).

\bibitem{gr1}
C. Coste, E. Falcon,  and S. Fauve, Phys. Rev. E \textbf{56}, 6104 (1997).

\bibitem{gr4} N. Boechler, G. Theocharis, S.  Job, P. G.  Kevrekidis, M. A. Porter, and  C. Daraio, Phys. Rev. Lett. \textbf{104}, 244302 (2010).

\bibitem{tunable} N. Boechler, J. Yang, G. Theocharis, P.  Kevrekidis, C. Daraio, Journal of Applied Physics, \textbf{109}, 7, 074906, (2011).

\bibitem{gr5} J. Hong, Phys. Rev. Lett. \textbf{94}, 108001 (2005).

\bibitem{Spadoni} A. Spadoni and C. Daraio, Proc. Natl. Acad. Sci. U.S.A, {\bf 107}, 7230, (2010).

\bibitem{Nature11} N. Boechler, G. Theocharis, and C. Daraio,  Nat. Mat. \textbf{10}, 665 (2011).

\bibitem{Gonella} R. Ganesh and S. Gonella, Phys. Rev. Lett. \textbf{114}, 054302 (2015).

\bibitem{PRL_Aurel} A. Merkel, V. Tournat, V. Gusev, Phys. Rev. Lett. {\bf 107}, 225502 (2011).

\bibitem{NJP_VT} V. Tournat, I. P\`erez-Arjona, A. Merkel, V. Sanchez-Morcillo, V. Gusev, New Journal of Physics {\bf 13}, 073042 (2011).

\bibitem{Nick} M. Hiraiwa, M. Abi Ghanem, S. P. Wallen, A. Khanolkar, A. A. Maznev, and N. Boechler,  arXiv:1510.05975.

\bibitem{PRL_Torsional} J. Cabaret, P. B\'equin, G. Theocharis, V. Andreev, V.E. Gusev, V. Tournat, Phys. Rev. Lett. {\bf 115}, 054301 (2015).

\bibitem{magnets} M. Moleron, A. Leonard and C. Daraio, J. Appl. Phys. {\bf 115}, 184901 (2014); J. Boisson, C. Rouby, J. Lee and O. Doar\'e,  EPL,  \textbf{109}, 34002, (2015); L. M. Nash, D. Kleckner, A. Read, V. Vitelli, A. M. Turner, and W. T. M. Irvine, Proc. Natl. Acad. Sci., 112, (47), 14495-14500, (2015); F. J. Sierra-Valdez, F. Pacheco-V\'azquez, O. Carvente, F. Malloggi, J. Cruz-Damas, R. Rechtman and J. C. Ruiz-Su\'arez, Phys. Rev. E,  \textbf{81}, 011301, (2010).

\bibitem{fujita} T. Fujita and M. Mamiya, J. Magn. Magn. Mater. {\bf 65}, 207 (1987).

\bibitem{rahmatian} A. Mehdizadeh, R. Mei, J. F. Klausner, N. Rahmatian, Acta Mech Sin {\bf 26}, 921 (2010).

\bibitem{Sm} See supplemental material at [URL] for the elements of the dynamical matrix {\bf S}.

\bibitem{Johnson} L.M. Schwartz, D.L Johnson and S. Feng, Phys. Rev. Lett. {\bf 52}, 831 (1984).

\bibitem{Pichard_PRE2014} H. Pichard, A. Duclos, J.-P. Groby, V. Tournat, V.E. Gusev, Physical Review E, 89(1), 013201 (2014).

\bibitem{Lydon} J. Lydon, G. Theocharis, C. Daraio, Phys. Rev. E,  {\bf 91}, 023208, (2015).

\end{thebibliography}

\end{document}